# Within-Journal Self-citations and the Pinski-Narin Influence Weights

Gangan Prathap[a] and Loet Leydesdorff[b]

[a]*A P J Abdul Kalam Technological University, Thiruvananthapuram, Kerala, India 695016*
e-mail: gangan_prathap@hotmail.com

[b]*Amsterdam School of Communication Research (ASCoR), University of Amsterdam, PO Box 15793, 1001 NG Amsterdam, The Netherlands; loet@leydesdorff.net*

**Abstract**

The Journal Impact Factor (JIF) is linearly sensitive to self-citations because each self-citation adds to the numerator, whereas the denominator is not affected. Pinski & Narin (1976) derived the Influence Weight (IW) as an alternative to Garfield's JIF. Whereas the JIF is based on raw citation counts normalized by the number of publications, IWs are based on the eigenvectors in the matrix of aggregated journal-journal citations without a reference to size: the cited and citing sides are combined by a matrix approach. IWs emerge as a vector after recursive iteration of the normalized matrix. Before recursion, IW is a (vector-based) non-network indicator of impact, but after recursion (i.e. repeated improvement by iteration), IWs can be considered a network measure of prestige among the journals in the (sub)graph as a representation of a field of science. As a consequence (not intended by Pinski & Narin in 1976), the self-citations are integrated at the field level and no longer disturb the analysis as outliers. In our opinion, this is a very desirable property of a measure of quality or impact. As illustrations, we use data of journal citation matrices already studied in the literature. Routines for the computation of IWs are made available at http://www.leydesdorff.net/iw.

**Introduction**

Cason and Lubotsky (1936) were the first authors to report that aggregated journal-to-journal cross-citations can be used to measure the *influence* of one journal on another. The objective of these authors was the quantitative measurement in terms of aggregated journal-journal citations of the extent to which each psychology journal influences and is influenced by other psychological journals. The journals are then considered as a proxy of fields. A similar journal-journal citation matrix in psychology was used by Daniel and Louttit (1953) to measure the similarity of the citation patterns of journals. These authors furthermore developed a first clustering of scientific journals. Kessler (1964) formulated a journal cross-citing matrix for physics journals and argued that specific types of information can be deduced from this matrix. Xhignesse and Osgood (1967) extended network-theory concepts to portray the relationships between journals and to measure their referencing similarities.

A problem in these matrices had remained the outlier on the main diagonal representing the within-journal self-citations. In two contributions, Price (1981a) and Noma (1982) proposed normalization procedures for these diagonal values. They noted that square matrices are common to the measurement of science, books, money, etc. These matrices register transactions between the members of a group. Price (1981b) argues that a set of five separate measures can be extracted from a given transaction matrix indicating size, quality, and self-interest in the cited and citing directions.

Earlier, Pinski and Narin (1976) proposed an iterative algorithm based on a matrix approach. From this perspective, the outlier is a characteristic organizing a subgraph of the matrix. Thus, these authors shifted the focus from the observable (raw) citation counts to what these counts mean in the context of the citation matrix under study. Different from citations as streams among individual journals, the subgraph among the journals can be considered as a representation of a scientific field.

Operationally Pinski & Narin (1976) first normalized the citation matrix and then an eigenvalue operation is used so that instead of a raw count of citations *C*, a (recursively) weighted count is generated that operationalizes the "prestige" of the citing journal in the field represented by the (sub)graph. This same idea was later reinvented by Page & Brin (Page *et al.*, 1998) as PageRank, the ranking algorithm of Google. The basic idea is that it matters who is citing: a more highly-cited citing agent is weighted as more important than a lower-cited one. From this perspective, citation values in the cells of a citation matrix are no longer considered as independent observations, but as recursively related outcomes of underlying processes.

Taking into account the "prestige" of the citing journal in a matrix from which a citation arises as a network measure, requires this iteratively recursive computation (Pinski and Narin 1976; Brin and Page 2001; Bergstrom 2007). In social network analysis, well-established tools (such as Pajek) allow for the computation of these recursive indicators.[1] From this perspective, the raw count of citations is a first non-network—since vector-based—measure. The "raw" count of citations can, for example, be considered as a measure of the "popularity" of the *journal* among other journals along the vector; recursively "weighted" counts of citations are assumed to measure "prestige" at the *field* level (Bollen *et al.* 2006; Yan and Ding 2010). Thus, the meaning of a citation is differently contextualized.

IWs were developed with the objective of providing an alternative to (and implicitly a critique of) the Journal Impact Factor (JIF). Garfield and Sher (1963) first measured the size of a journal by the count of all articles $P$ published in the journal during a chosen window (called the publications window). This $P$ is a measure of journal performance. The output measure is the number of citations $C$ received by these $P$ articles from all articles published in the other journals in the network during a specified period called the citation window. From these, one can derive a proxy of quality called impact $i = C/P$. In the case of the Journal Impact Factor (JIF), the numerator $C$ is number of references in the current year (citation window) to the articles published in the previous two years (publications window) while the denominator is the number of articles $P$ published during the same period. JIF is meant to be a ratio of values (Garfield, 1972; cf. Prathap 2018).

Within-journal self-citations can affect a JIF dramatically. Fassoulaki *et al.* (2000), for example, studied self-citations in the 1995 and 1996 issues of six anaesthesia journals by calculating the self-citing and self-cited rates for each journal. Among these six journals, the journal *Anesthesiology* had the highest self-citing rate (57%), and also the highest self-cited rate (35%). Self-citations thus increased the count of citations by 35% (the numerator in the formula for the JIF) while the denominator, which is the count of publications remained unchanged. Consequently, the JIF of *Anaesthesiology* is driven for 35% by self-citations. A correction for self-organization changes the ranking of journals in most fields.

---

[1] Prathap *et al.* (2016), for example, used a "tournament" metaphor that was introduced by Ramanujacharyulu (1964) for defining a new dimensionless network property.

In the following sections of this paper we focus on the Pinski-Narin Influence Weights as dimensionless metrics for journal evaluation that arise naturally from a network approach without a reference to size as operationalized by the number of papers *P*. To illustrate the behavior of the indicator, we develop the argument using the journal citation matrix among eight leading biochemistry journals published by Price (1981b, at p. 59).

**Price's example of a subfield of biochemistry journals**

Frandsen (2007: 48) illustrated the basic concepts using the simplified citation matrix in Table 1. One can depict an ecosystem or sub-graph built around the cluster of other journals O closest to a Journal J under study. Row-wise, one lists the citations that are made to each entity from the others ("cited") and column-wise the references each entity makes ("citing") are counted. Thus, the within-journal citations (and within-journal references) are the diagonal terms *S* and *X*, and the cross-terms are the journal-to-journal citations and references respectively. That is, Journal J cites itself *S* times, is cited *d* times by the other journals and is citing the other journals *g* times. The cited-citing ratio of Journal J with self-citations is then *(S+d)/(S+g)*, and that without self-citations is *d/g*.

Table 2 provides the matrix of citation relations among eight biochemistry journals (Price 1981b): **Z = [Z$_{ij}$]**. Many properties of such matrices are known. Among other things, the matrix can be multiplied by itself. This can recursively be repeated indefinitely so that the $k^{th}$ power of the matrix is **Z$^k$**. This matrix multiplication can be done, for example, in Excel using the function MMult(). However, this function is limited in Excel to 73 * 73 arrays. Larger matrices can be handled, for example, in Pajek, or by the routine "power.exe" available at http://www.leydesdorff.iw (limited to matrices of 1024*1024 rows and columns).

Whereas matrix multiplication (by itself) is symmetrical along the row and column dimensions, multiplication by the probability vector $p(k)$ can be expected to converge and result in a vector containing the effectively weighted values of total citations, but asymmetrically for "cited" and "citing." Differently from Ramanujacharyulu (1964), Pinski & Narin (1976) did not depart from the eigenvectors of the two matrices, namely the cited and citing forms (where the latter is the transpose of the former). Whereas the computation of the JIF proceeds by normalizing citations by the number of publications, Pinski & Narin (1976) proposed to normalize first the *citations* of a journal by dividing by the aggregated total number of ("citing") *references* along the column vector; and vice versa for the normalization in the citing dimension. The advantage of this normalization is that one divides among units with the same dimension and the result is therefore dimensionless. Table 3 shows the normalized reconstruction of the original matrix in Table 2; Table 4 shows the convergence of the vector $p(k)$ in the case of the biochemistry journal ecosystem (subgraph in the graph theoretical sense) for the IW indicator after recursion (i.e. 7 iterations) for the cases with and without self-citations. Note that with sufficiently large iteration, it will go asymptotically to zero.

A formal elaboration of this procedure was provided by Todeschini, Grisoni, and Nembri (2015, p. 330). The recursive procedure for formalizing the computation of $p_i(k)$ is given in graph-theoretical terms by Ramanujacharyulu (1964). An algorithmic implementation using the so-called Stodola method of iteration is provided by Dong (1977). In the appendix, we provide an Excel procedure for calculating IW from a citation matrix. The corresponding Excel file is available for download at http://www.leydesdorff.net/iw/price.xlsx . A disadvantage of Excel is the limitation to 73 rows and columns. A general purpose program for the computation of influence weights on the basis of square matrices "vector.exe" is therefore provided at http://www.leydesdorff.net/iw/index.htm . Vector.exe is limited to 1024*1024 rows and columns.

Figure 1 shows two ways in which the IW indicators for the eight bio-chemistry journals (Price 1981b) can be displayed: before and after recursion and with or without self-citations. The IW indicator after recursion, is virtually insensitive to self-citation. The intercept of the trendline (in Excel) is 0.00, the slope is 1.00, and the correlation is 1.00. This indicates that the linear trend line through the data points passes through the origin, has a unit slope, and a goodness of fit very close to 1.0. For example, the JIF of the *Journal of Biological Chemistry* in the matrix of eight biochemistry journals changes by 34% when the 9,384 within-journal self-citations in Table 3 are subtracted from the margin total of 27,596 citations. However, the change in IW after recursion is only 0.14% after 7 iterations, and diminishes exponentially with iteration number!

**Concluding remarks**

IW is a dimensionless indicator that emerges from the graph-theoretic properties of the citation network. When IW converges, it can be considered as a proxy for the quality of the journal's performance *in the relevant network* (Price 1981b; Pinski & Narin 1976). In this study, we have seen that after recursion (i.e. repeated improvement by iteration of the matrix multiplication), the IW is remarkably insensitive to surplus self-citations. This is a very desirable property of a measure of quality.

Self-citations highly exaggerate the Journal Impact Factor, the inflation being linear with self-citation. Price's (1981a) normalization improved the indicator of how journals perform within a journal ecosystem. Pinski & Narin's (1981) IWs are insensitive to self-citation. It can be formally proved that the recursive iteration lets IWs to converge to the dominant eigenvector and that the converged IWs are insensitive to self-citations (i.e. the diagonal terms) and this is being communicated separately (Mukherjee & Prathap 2019).

Furthermore, one can specify the differences between IWs and Ramanujacharyulu's (1964) Power-Weakness Ratio (PWR). In the PWR approach, the eigencomputation is performed separately on the cited and citing dimensions of the matrices and then the ratio is taken of the resulting vectors (Prathap 2019). In the IW approach, the matrix is first normalized and the eigencomputation is performed on this matrix. However, both PWR and IW should be used only with homogeneous sets (Leydesdorff *et al.*, 2016).

**Acknowledgment**

We thank Inga Ivanova and two anonymous referees for comments on an earlier draft. We are grateful to ISI/Clarivate ™ for providing us with JCR data.

**Appendix 1: The computation of Influence Weights**

**A. Excel**

1) A file is provided at http://www.leydesdorff.net/iw/price.xlsx. containing Price's (1981b) 8 × 8 cited-citing matrix and the normalized matrix in the first two sheets respectively and repeated in array (C3:J10) of the third sheet labelled 'w sc' (that is, "with self-citations"). The matrix for the case of zero within-journal self-citations is found in the fourth sheet labelled 'wo sc' ("without self-citations") in this same sheet.

2) In sheet 3 labelled 'w sc' the first matrix multiplication (using the mmult() function in Excel) multiplies each row of this matrix with the start vector (J3:J10), taken as a vector with each element having the value 1.This actually gives the raw count of citations, and is kept at (L3:L10). The new eigenvector is obtained at column M by normalizing this so that it becomes a stochastic vector. The multiplication is then done repeatedly. At the end of the $k^{th}$ cycle one obtains the vector *p(k)*.

3) The iteration can be repeated with the transposed matrix. One obtains the vector *q(k)* at the end of the $k^{th}$ cycle.

4) Ramanujacharyulu's (1964) power-weakness ratio *r* is then given by *r(k) = p(k)/q(k)* at the end of the *k* cycles.

5) The recursion is repeated with the normalized **Z** matrix in order to obtain the IW vector. Note that at *k*=1, the PWR and IW values are exactly the same, as expected.

6) The MMULT function returns #VALUE! if the output exceeds 5460 cells (n ≤ 73); see at https://support.microsoft.com/kb/166342?wa=wsignin1.0. In that case, use option B below.

**B. Using Vectors.exe and Power.exe at http://www.leydesdorff.net/iw**

1) Export the transaction matrix as comma-separated variables file to text.csv. The file should be "pure ASCII"; that is, MS-DOS with Carriage Return and Line Feed (CR + LF) at the end of each line. (Use WordPad or Edit++.) The file should not contain a first line with headings; the file name "text.csv" is obligatory.

2) "text.csv" can be read by vector.exe to be downloaded first and stored in the same folder.

3) Output of vector.exe with the possibly converging vector for 15 iterations.

4) The file narin1.dbf contains the normalized data file before iteration; both in the transposed direction and before this in the non-transposed one. Files are overwritten in subsequent runs.
5) One can replace "text.csv" by a file of this name but containing the transposed for the "citing" analysis.
6) "Narin1.dbf" can be used for making another (normalized) version of text.csv. This file can be exported from Excel, SPSS, etc. The normalized file can also be input into power.exe in order to make higher-order power matrices.

**Table 1**. Citation matrix of an ecosystem or sub-graph built around the cluster of journals O closest to Journal J.

| Citation matrix | | Citing | | Citations |
|---|---|---|---|---|
| | | Journal J | Others | |
| Cited | Journal J | $S$ | $d$ | $S+d$ |
| | Others | $g$ | $X$ | $g+X$ |
| References | | $S+g$ | $d+X$ | $S+d+g+X$ |

Table 2. The **Z** matrix of the cross citing terms among the eight bio-chemistry journals as a subgraph of the main graph of all the journals listed in the 1977 Journal Citation Index (source: Price 1981b).

|  | Bio-chemistry journals | Citing | | | | | | | | CITATIONS |
|---|---|---|---|---|---|---|---|---|---|---|
| **Cited** | **J. Biol. Chem** | 9384 | 6181 | 2107 | 3750 | 609 | 2335 | 719 | 2511 | 27596 |
|  | **Bio. Bio. Aeta** | 2406 | 7550 | 865 | 1757 | 365 | 1478 | 408 | 1120 | 15949 |
|  | **Proc. N.A.S.** | 2770 | 2184 | 3995 | 1946 | 1470 | 488 | 1239 | 1329 | 15421 |
|  | **Biochem. U.S.** | 2553 | 2591 | 1057 | 3827 | 299 | 653 | 601 | 887 | 12468 |
|  | **Nature** | 1007 | 1230 | 1407 | 837 | 2963 | 379 | 603 | 630 | 9056 |
|  | **Biochem. J.** | 1183 | 1812 | 326 | 632 | 201 | 2464 | 150 | 528 | 7296 |
|  | **J. Mol. Bio.** | 1109 | 1136 | 1251 | 1347 | 504 | 216 | 2545 | 367 | 8475 |
|  | **Bio. Bio. R.C.** | 1624 | 1719 | 695 | 1040 | 263 | 564 | 241 | 1313 | 7459 |
| **REFERENCES** |  | 22036 | 24403 | 11703 | 15136 | 6674 | 8577 | 6506 | 8685 | 103720 |

Table 3. The normalized **Z** matrix of the cross citing among the the eight bio-chemistry journals as a subgraph of the main graph of all the journals listed in the 1977 Journal Citation Index (Price 1981b) after Pinski-Narin recursion (1976).

| Normalized Matrix | Bio-chemistry journals | Citing | | | | | | | | CITATIONS |
|---|---|---|---|---|---|---|---|---|---|---|
| Cited | J. Biol. Chem | 0.426 | 0.280 | 0.096 | 0.170 | 0.028 | 0.106 | 0.033 | 0.114 | 1.252 |
| | Bio. Bio. Aeta | 0.099 | 0.309 | 0.035 | 0.072 | 0.015 | 0.061 | 0.017 | 0.046 | 0.654 |
| | Proc. N.A.S. | 0.237 | 0.187 | 0.341 | 0.166 | 0.126 | 0.042 | 0.106 | 0.114 | 1.318 |
| | Biochem. U.S. | 0.169 | 0.171 | 0.070 | 0.253 | 0.020 | 0.043 | 0.040 | 0.059 | 0.824 |
| | Nature | 0.151 | 0.184 | 0.211 | 0.125 | 0.444 | 0.057 | 0.090 | 0.094 | 1.357 |
| | Biochem. J. | 0.138 | 0.211 | 0.038 | 0.074 | 0.023 | 0.287 | 0.017 | 0.062 | 0.851 |
| | J. Mol. Bio. | 0.170 | 0.175 | 0.192 | 0.207 | 0.077 | 0.033 | 0.391 | 0.056 | 1.303 |
| | Bio. Bio. R.C. | 0.187 | 0.198 | 0.080 | 0.120 | 0.030 | 0.065 | 0.028 | 0.151 | 0.859 |
| **REFERENCES** | | 1.576 | 1.716 | 1.063 | 1.187 | 0.763 | 0.694 | 0.722 | 0.696 | 8.416 |

Table 4. Summary of the biochemistry journal ecosystem (subgraph in the graph theoretical sense) and the IW indicator after recursion (i.e. 7 iterations) for the cases with and without self-citations.

| Subgraph | Journal name | With self-citations | Without self-citations | % Change |
|---|---|---|---|---|
| Biochemistry journals (Price 1981b) | The Journal of Biological Chemistry | 0.1363 | 0.1361 | -0.14 |
| | Biochimica et Biophysica Acta | 0.0592 | 0.0591 | -0.17 |
| | Proceedings of the National Academy of Sciences of the United States | 0.1739 | 0.1740 | 0.04 |
| | Biochemistry | 0.0870 | 0.0869 | -0.12 |
| | Nature | 0.1942 | 0.1947 | 0.21 |
| | Biochemical Journal | 0.0809 | 0.0807 | -0.20 |
| | Journal of Molecular Biology | 0.1770 | 0.1772 | 0.11 |
| | Biochemical and Biophysical Research Communications | 0.0914 | 0.0913 | -0.12 |

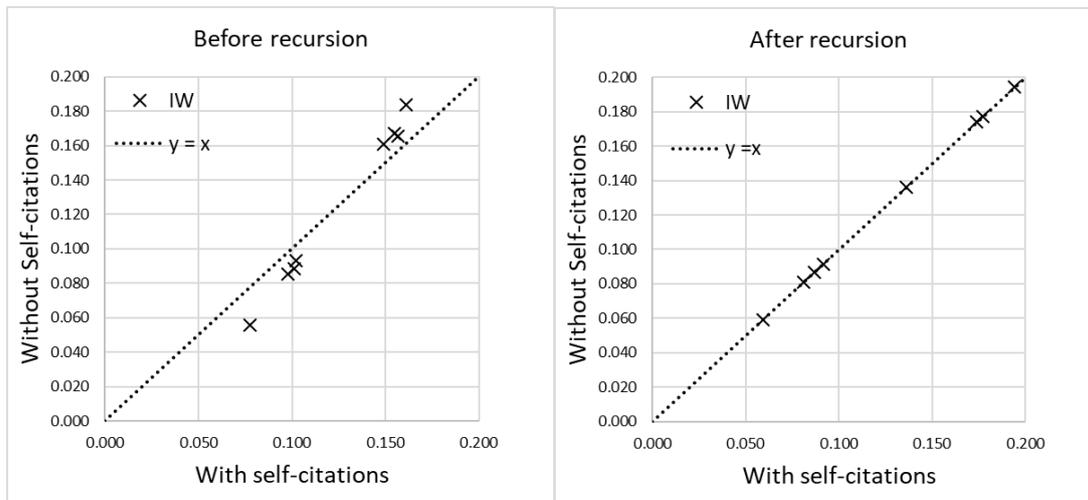

Figure 1. The IW indicators for the eight bio-chemistry (Price 1981b) before and after recursion, and with and without self-citations.